\documentclass[pra,preprint,floatfix,onecolumn,superscriptaddress]{revtex4-1}
\usepackage[dvips]{graphicx}
\usepackage[english]{babel}
\usepackage{amsmath}
\usepackage{amssymb}
\usepackage{tensor}

\newcommand{\vk}{\mathbf{k}}

\newcommand{\be}{\begin{eqnarray}}
\newcommand{\ee}{\end{eqnarray}}
\newcommand{\p}{\partial}

\newcommand{\da}{a^{\dagger}}

\def\ket#1{|#1\rangle}
\def\bra#1{\langle #1 |}
\def\ep#1{\langle #1 \rangle}

\begin{document}

\title{Geometric phases of Topological Systems under Quench Process}

\author{Zhan Gao}
\affiliation{College of Physics, Sichuan University, Chengdu, Sichuan 610064, China}

\author{Yan He}
\affiliation{College of Physics, Sichuan University, Chengdu, Sichuan 610064, China}
\email{heyan$_$ctp@scu.edu.cn}

\begin{abstract}
We study the time evolution of geometric phases of one dimensional topological models under the quench dynamics.
Taking the Creutz ladder model as an example, it is found that the Berry phase is fixed as the parameter is suddenly tuned across the topological phase boundary, given that the chiral symmetry of the model is preserved. At finite temperature, the Uhlmann phase displays abrupt jumps between the two quantized values, which indicates the topological transition at certain times after the quench. Both the Berry and Uhlmann phase will deviate from quantized values if the chiral symmetry of the model is broken. The relation between the Uhlmann phase and Loshmidt rate function under the quench process is also discussed.

\end{abstract}

\maketitle

\section{Introduction}

In recent years, with the development of quantum technology, the dynamics of the quench process of closed quantum systems has attracted a lot of attentions of both theoretical and experimental studies \cite{Bloch,Polkovnikov,Gogolin}. The quench processes of many quantum systems are frequently associated with the so-called dynamical quantum phase transition \cite{Heyl-13,Heyl-review}. These transitions can be detected by the zeros of the Loschmidt amplitudes or the divergences of the Loschmidt rate functions, which is in analogy to the non-analytical behaviors of thermodynamic phase transitions \cite{Heyl-14,Heyl-15,HouPRB20}. For quantum systems with nontrivial topology, other than the rate functions, one may also wonder about how the topological indices evolve under the quench dynamics. This is exactly the question we will try to address in this paper.

In one-dimensional systems, the geometric phases \cite{Geo-phase-book} provide the widely used indices to understand the topological properties of topological insulators and superconductors~\cite{Zhang_TIRev,Kane_TIRev}. At zero temperature, the central notion is the Berry phase which is the accumulation of Berry connection along a closed loop in the parameter space \cite{Berry,Zak89}. In some sense, the Berry phase reflects the twist of the phase factor of the eigenstates as one travels around this loop \cite{Nakahara}. For one-dimensional systems, the Berry phase is usually quantized to the values of $0$ and $\pi$, which corresponds to the even and odd winding number of the mapping defined by the Hamiltonian.

At finite temperatures, the physics of mixed states is completely described by their density matrices. There are many attempts to generalize the Berry connection to the case of mixed states \cite{Sjoqvist00,Diehl,Diehl18}. In this paper, we will follow the method proposed by Uhlmann \cite{Uhlmann,Uhlmann1,Uhlmann2} long time ago, which is closely parallel to the notion of Berry connection in the geometric sense. The key step in defining Uhlmann connection is to decompose the density matrices into amplitudes and then set up a parallel condition that is strong enough to uniquely determine the arbitrary phase factors of the amplitudes. Similar to the Berry phase, the Uhlmann phase is also the accumulation of Uhlmann connection along a closed loop in the parameter space. The Uhlmann phase has been successfully used as a topological index to understand many topological systems \cite{Viyuela14,Viyuela2,Huang14,Mera17,HeChern18,Zhang21} and spin systems \cite{Galindo21,HouPRA21} at finite temperature. In terms of the Uhlamnn phase, it is usually found that there is a critical temperature $T_c$ which separates the topologically non-trivial phase from the trivial one.

In this paper, we will take the Creutz ladder \cite{Creutz} as our example to investigate the time evolution of the Berry and Uhlmann phases under the quench process. For two-band models, the time evolution can be visualized as a closed loop on a unit sphere whose shape is deformed as time flows. From this point of view, we will show that the Berry phase will keep constant values after the quench, given the condition that the symmetry of the model is preserved. On the other hand, the Uhlmann phase will predict topological transitions at certain points of time after the quench. We will also try to compare the behavior of Uhlmann phase with that of Loschmidt rate functions.

The rest of this paper is organized as follows. In section \ref{sec-quench}, we provide a geometric description of the quench process for generic two-band models. Then in section \ref{sec-U}, we briefly review the concepts of Berry and Uhlmann phases which is used as topological indices. The detailed calculations and numerical results of these two types of geometric phases under the quench process will be presented in section \ref{sec-CL}. At last, a simple summary will be given in section \ref{sec-con}.

\section{Quench process for two-band models}
\label{sec-quench}

In this paper, we will be mainly concerned with the quench process of two-band models. Since the topological properties are only determined by the eigenstates, we can always shift the average energy of the two bands to be zero, which makes the Hamiltonian traceless.  Then the most generic two-band traceless Hamiltonian can be expanded by Pauli matrices $\sigma_i$ as
\be
H=\sum_{i=1}^3 R_i(\vk)\sigma_i
\ee
Clearly, the above Hamiltonian can be represented by this 3D vector $R_i$ whose components are functions of lattice momentum $\vk$.

Now the quench process can be described as follows. First, suppose the initial Hamiltonian is given by $H_0=R^0_i(\vk)\sigma_i$. Then we make a sudden change of the parameters, thus the system will evolve under another Hamiltonian $H_1=R^1_i(\vk)\sigma_i$ where $R^1_i\neq R^0_i$. Here we assume both $H_0$ and $H_1$ are time independent, thus the time evolution can be computed analytically.

At zero temperature, we assume that the initial state $\ket{\psi_0}$ is an eigenstate of $H_0$. After the quench process, it evolves into
\be
\ket{\psi(t)}=\exp(-i H_1 t)\ket{\psi_0}
\ee
Since $H_1$ does not commute with $H_0$,  the resulting $\ket{\psi(t)}$ is usually a superposition of the two eigenstates of $H_0$. In the next section, we will consider the time evolving of its Berry phase under the quench process.

At a finite temperature $T$, we assume that the initial state is a thermal equilibrium state with the following density matrix
\be
\rho(0)=\frac{1}{Z}e^{-H_0(\vk)/T},\qquad Z=\textrm{Tr}[e^{-H_0(\vk)/T}]
\ee
for a fixed momentum $\vk$.

Then the system is evolved under $H_1=\sum_i R^1_i\sigma_i$ where $R^1_i\neq R^0_i$. In the quench process, at some later time $t$, the density matrix can be computed as follows
\be
\rho(t)=e^{-iH_1t}\rho(0)e^{iH_1t}=\frac1Z\exp\Big(-\frac1T e^{-iH_1t}(R^0_i\sigma_i)e^{iH_1t}\Big)
\ee
Therefore, the time evolution is equivalent to making a unitary transformation of the initial Hamiltonian $H_0$.
According to the well-known relation between the spinor and vector representation of $SU(2)$ group, we have the following identity
\be
&&e^{-i\theta n_a\sigma_a/2}(\sigma_b r_b)e^{i\theta n_a\sigma_a/2}=\sigma_a R_{ab}r_b\\
&&R_{ab}=\delta_{ab}\cos\theta+n_an_b(1-\cos\theta)-\epsilon_{abc}n_c\sin\theta
\ee
Here $r_a$ and $n_a$ are 3D vectors and $n_a$ has unit length. The angle $\theta$ is a real number.
The matrix $R_{ab}$ rotates the vector $r_a$ around the axes along $n_a$ with angle $\theta$. To ease the notation, we assume that the repeated indices imply summations.

Making use of the above identity, the unitary transformation by $e^{-iH_1t}$ is equivalent to a rotation of the vector $R^0_i$
\be
&&R_a(t)=R^0_a\cos(2R^1 t)+\hat{R}^1_a(\hat{R}^1_b R^0_b)[1-\cos(2R^1 t)]-\epsilon_{abc}R^0_b\hat{R}^1_c\sin(2R^1 t)
\label{eq-Rotate}
\ee
with $R^1=\sqrt{R^1_a R^1_a}$ and $\hat{R}^1_a=R^1_a/R^1$.  Thus, the effect of the quench process is to rotate the initial vector $R^0_a$ around the axes along $\hat{R}^1_a$ with the angle $2R^1 t$ to obtain a new vector $R_a(t)$. With this new vector $R_a(t)$, $\rho(t)$ can be expressed as
\be
&&\rho(t)=\frac1Z\exp\Big(-\frac1T R_a(t)\sigma_a\Big)
\ee
Here the partition function $Z$ is the same as the initial state. Based on these density matrices, Uhlmann connections can be computed for any later time $t$.

\section{A Brief Review of Berry and Uhlmann Connections}
\label{sec-U}

We want to consider the behavior of geometric phases under quench processes. At zero temperature, the most widely used geometric phases are the Berry phases. Let us consider a given energy band of a lattice system. Then the eigenstate of this band $\ket{u_n(\vk)}$ depends on the parameter of lattice momentum $\vk$. For this band, the Berry connection is defined as
\be
A_\mu=-i\ep{u_n(\vk)|\frac{\p}{\p k_\mu}|u_n(\vk)}.
\ee
Here the Berry connection measures the phase difference of the neighboring wave functions in the $\vk$ space. The Berry phase is defined to be the phase accumulated as one travels around a closed loop in the $\vk$ space, which can be written as
\be
\Phi=\oint_C A_\mu d k_\mu
\ee
Here $C$ denotes a closed loop in $\vk$ space. Although $A_\mu$ is not gauge invariant, it is easy to see that the above defined $\Phi$ is gauge invariant and thus is also an observable.

At finite temperatures, the above notion of Berry connection can be generalized to the so-called Uhlmann connection \cite{Uhlmann,Uhlmann1,Uhlmann2}.
For a mixed state, the physical properties are determined by the density matrix $\rho$ instead of wave functions. It is convenient to make a spectral decomposition of density matrices as $\rho=\sum_i p_i\ket{u_i}\bra{u_i}$. Here  $\ket{u_i}$ for $i=1,\cdots,n$ are eigenstates and $p_i$ is the Boltzmann weight for systems at thermal equilibrium. The concept of Uhlmann connection is based on the amplitude decomposition of the density matrix as
\be
\rho=ww^{\dagger},\qquad w=\sqrt{\rho}\,U.
\ee
Here $w$ can be thought of as the counterpart of the wave function for the mixed states. Note that for a fixed $\rho$, $w$ can be multiplied by an arbitrary unitary matrix $U$ but still reproduce the same $\rho$. This matrix $U$ represents a $U(n)$ phase factor, which is very similar to the $U(1)$ phase factor of a wave function.

In order to determine a preferred phase factor $U$, Uhlmann \cite{Uhlmann} proposed the following parallel condition:
\be
w_1^{\dagger}w_2=w_2^{\dagger}w_1=C>0.
\ee
Where $C>0$ means that $C$ is a Hermitian and positive definite matrix. With this condition, the relative phase factor between $w_1$ and $w_2$ is uniquely determined.
Let us see how to derive this as follows.

Consider two different density matrices and their amplitudes $w_1=\sqrt{\rho_1}U_1$ and $w_2=\sqrt{\rho_2}U_2$. The self multiplication of the parallel condition gives
\be
C^2=w_1^{\dagger}w_2w_2^{\dagger}w_1=U_1^{\dagger}\sqrt{\rho_1}\rho_2\sqrt{\rho_1}U_1
\ee
After taking the square root, one finds that
\be
C=U_1^{\dagger}\sqrt{\sqrt{\rho_1}\rho_2\sqrt{\rho_1}}\,U_1,
\ee
Substitute $C$ back to the parallel condition, we find a relative phase factor as
\be
U_2U_1^{\dagger}=\sqrt{\rho_2^{-1}}\sqrt{\rho_1^{-1}}\sqrt{\sqrt{\rho_1}\rho_2\sqrt{\rho_1}}.
\label{eq-UU}
\ee
This result can be considered as a finite version of the Uhlmann connection. The above formula requires the density matrix $\rho$ to be a full-rank matrix, since the inverse of $\rho$ must be inserted.

To simplify the above result, one can also work with an infinitesimal Uhlmann connection. To this end, we can set $\rho_1=\rho(\vk)$ and $\rho_2=\rho(\vk+\Delta\vk)$ in Eq.(\ref{eq-UU}), then the infinitesimal Uhlmann connection can be defined as
\be
 A^U_\mu=\p_\mu U U^{\dagger},
\ee
where $\p_\mu=\frac{\p}{\p k_\mu}$. After some straightforward calculations, the explicit result of the Uhlmann connection can be written as
\be
A^U_\mu&=&\ket{u_i}\bra{u_i}\frac{[\p_\mu\sqrt{\rho},\,\sqrt{\rho}]}{p_i+p_j}\ket{u_j}\bra{u_j}\nonumber \\
&=&\frac{(\sqrt{p_i}-\sqrt{p_j})^2}{p_i+p_j}\ket{u_i}\bra{u_i}\p_\mu\ket{u_j}\bra{u_j}.
\label{eq-AU}
\ee
We would like to mention that, if one makes another gauge choice, then $A^U$ will transform like an ordinary $U(n)$ non-Abelian gauge field. Similar to the Berry connection, we can construct a gauge invariant Wilson loop based on the Uhlmann connection as
\be
V=\mathcal{P}\exp\Big(\oint_C A^U_{\mu}d k_{\mu}\Big).
\ee
Here $C$ represents a closed loop in the $\vk$ space and $\mathcal{P}$ represents the path ordering. Now suppose we pick a starting point at the loop $C$. The amplitude at this point is given by $w_1=\sqrt{\rho_0}$. If we parallel transport $w_1$ along the loop $C$ and back to starting point, the amplitude becomes $w_2=\sqrt{\rho_0} V$ which is usually different from $w_1$. Now we can define the Uhlmann phase as the phase angle of the overlap between $w_1$ and $w_2$, which is given as follows
\be
\Phi^U=\arg\textrm{Tr}(w_1^{\dag}w_2)=\arg\textrm{Tr}(\rho_0 V).
\ee
This index has already been used to study several 1D topological systems at finite temperatures~\cite{Viyuela14}.

\section{Examples of geometric phases under quench process}
\label{sec-CL}

\begin{figure}
\centering
\includegraphics[width=0.7\columnwidth]{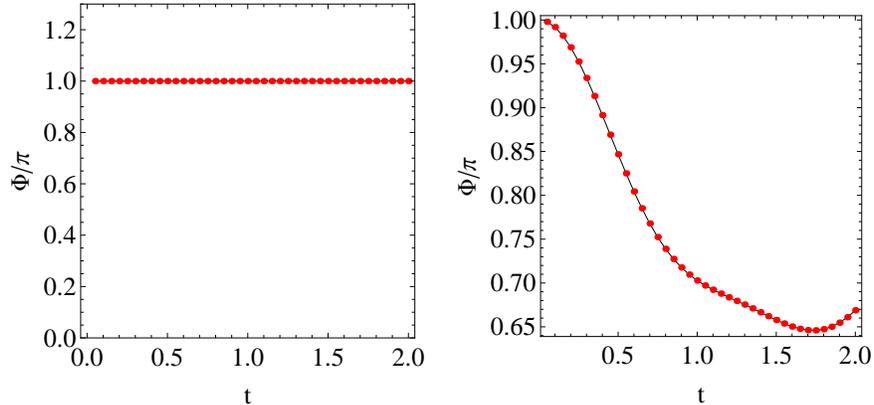}
\caption{The Berry phase $\Phi$ as a function of time for Creutz ladder of Eq.(\ref{eq-CL}) (left panel) and Eq.(\ref{eq-CLp}) (right panel). In the left panel, $m$ changes from $0.6$ to $1.6$. In the right panel, $(m,\gamma)$ changes from $(0.6,0)$ to $(1.6,0.5)$. $\theta=0.3\pi$ for both cases.}
\label{fig-Berry}
\end{figure}

As a specific example, we consider the quench process of the Creutz ladder, which is a typical one-dimensional topological model. In real space, the Hamiltonian of this model can be written as
\be
H=\sum_j\Big[\frac w2(e^{i\theta}\da_j a_{j+1}+e^{-i\theta}b^{\dag}_j b_{j+1})
+\frac w2(\da_j b_{j+1}+b^{\dag}_j a_{j+1})+m \da_j b_j+\textrm{h.c.}\Big]
\ee
Here $a$ and $b$ are fermion annihilation operators for A and B sublattices. $m$ is the hopping constant inside each unit cell. $w$ is the hopping constant between neighboring unit cells and we set $w=1$ as an energy unit. The phase angle $\theta$ represents the magnetic flux applied to each unit cell.

If we transfer the above Hamiltonian to momentum space and ignore the term proportional to identity matrix, we find that
\be
H=\sum_i R_i\sigma_i,\quad
R_i=(m+\cos k,\,0,\,\sin\theta\sin k)
\label{eq-CL}
\ee
In this form, it is easy to see that $\sigma_2 H \sigma_2=-H$. Therefore, the Creutz ladder has chiral symmetry and belongs to the AIII symmetry class \cite{Chiu2016}. As a comparison example, we also introduce a modified Creutz ladder as
\be
H'=\sum_i R_i\sigma_i,\quad
R_i=(m+\cos k,\,\gamma,\,\sin\theta\sin k)
\label{eq-CLp}
\ee
with a constant $\gamma\neq0$. The $\gamma\sigma_2$ term breaks the chrial symmetry of the original model.

At $T=0$, the topological properties of Creutz ladder can be easily understood by the vector $R_i(k)$. As $k$ increases from $0$ to $2\pi$, $R_i(k)$ sweeps an ellipse on the $xz$ plane. If $|m|<1$, this ellipse encloses the coordinate origin, which means the winding number is one. On the other hand, if $|m|>1$, then the coordinate origin is outside the ellipse and the winding number is zero. Therefore, $m=\pm1$ is the boundary between the topological and trivial phases.

\subsection{Berry phase}

We first consider the time-evolution of the Berry phase under quench process. In Figure \ref{fig-Berry}, we plot the Berry phase $\Phi$ as the function of time. In the left panel, we consider the Creutz ladder of Eq.(\ref{eq-CL}), which is initially set with $m=0.6$. Then we suddenly change $m$ to $m=1.6$. Although we tune $m$ across the phase boundary between the topological non-trivial and trivial phases, we find that $\Phi$ stays at the same non-trivial value $\pi$ for the whole time. If we start from an initial state with $|m|>1$, $\Phi$ will always take zero value. This constant behavior is due to the symmetry of Creutz ladder, as we will explain in detail later.

To break this symmetry, we consider the modified Creutz ladder of Eq.(\ref{eq-CLp}). In the right panel of Figure \ref{fig-Berry}, we show the time evolution of $\Phi$ under the quench from $(m,\gamma)=(0.6,0)$ to $(1.6,0.5)$. One can see that the Berry phase deviates from the quantized value $\pi$. Since the chiral symmetry is broken in the modified Creutz ladder, there is no topological index in this model any more.

Now we want to explain why $\Phi$ of the Creutz ladder always stays as a constant under the quench process. The initial Hamiltonian $H_0=R^0_i\sigma_i$ can be represented by a loop traced out by the vectors of $R^0_i(k)$ with $k\in(-\pi,\pi)$. The time evolution under the Hamiltonian $H_1=R^1_i\sigma_i$ is equivalent to a rotation around vector $R^1_i(k)$. Making use of Eq.(\ref{eq-Rotate}), the resulting vector is $R_i(k,t)$ at a fixed time $t$. We can treat the Berry connection as a vector field in $R_i$ space, then the Berry phase can be transferred to a surface integral by the Stoke's theorem as \cite{Qi2}
\be
&&\Phi=\int_0^{2\pi}A(k)dk=\int_C A_a\,d R_a\nonumber\\
&&=\frac12\iint_S\frac{\epsilon_{abc}R_a dR_b\wedge dR_c}{2R^3}
=\frac12\iint_S\frac12\epsilon_{abc}\hat{R}_a d\hat{R}_b\wedge d\hat{R}_c
\ee
Here we have defined $R=|{\bf R}|$ and unit vector $\hat{R}_i=R_i/R$. The last surface integral calculates the area of the region $S$ which is bounded by the curve of $\hat{R}_i(k)$ on the unit sphere. In the following, we will see that this integral can be determined solely by a symmetry property of the $\hat{R}_i(k)$ curve.

First, note that $R^0_i(k)$ describes a curve on the $xz$ plane, which is symmetric about the $x$-axis
\be
R^0_1(k)=R^0_1(-k),\quad
R^0_3(k)=-R^0_3(-k)
\ee
Similarly, the curve of $R^1_i(k)$ corresponding to $H_1$ is also symmetric about the $x$-axes
\be
R^1_1(k)=R^1_1(-k),\quad
R^1_3(k)=-R^1_3(-k)
\ee
According to the rotation formula Eq.(\ref{eq-Rotate}), the time evolved vectors $\hat{R}_i(k)$ will satisfy the following relations
\be
\hat{R}_1(k)=\hat{R}_1(-k),\quad \hat{R}_2(k)=-\hat{R}_2(-k),\quad
\hat{R}_3(k)=-\hat{R}_3(-k)
\ee
If we parameterize the vector $\hat{R}_i$ by spherical coordinate $(\theta,\phi)$, then the curve of $\hat{R}_i(k)$ can be represented by a function $\theta_0(\phi)$ which satisfies
\be
\theta_0(-\phi)=\pi-\theta_0(\phi)
\label{eq-t0}
\ee
Then the surface area can be computed as
\be
&&\textrm{area}
=\int_0^\pi d\phi\int_0^{\theta_0(\phi)}\sin\theta d\theta
+\int_{-\pi}^0 d\phi\int_0^{\theta_0(\phi)}\sin\theta d\theta\nonumber\\
&&=\int_0^\pi d\phi\int_0^{\theta_0(\phi)}\sin\theta d\theta
+\int_0^\pi d\phi\int_0^{\pi-\theta_0(\phi)}\sin\theta d\theta=2\pi
\ee
Then it is easy to see that $\Phi=\frac12$area$=\pi$. The above calculation does not depend on the function form of $\theta_0(\phi)$. The symmetry property of Eq.(\ref{eq-t0}) is enough to guarantee that the Berry phase $\Phi=\pi$.
In the modified Creutz ladder, the extra $\gamma\sigma_2$ term breaks this symmetry, thus the Berry phase is not a constant any more in this case.

\subsection{Uhlmann phase}

\begin{figure}
\centering
\includegraphics[width=0.7\columnwidth]{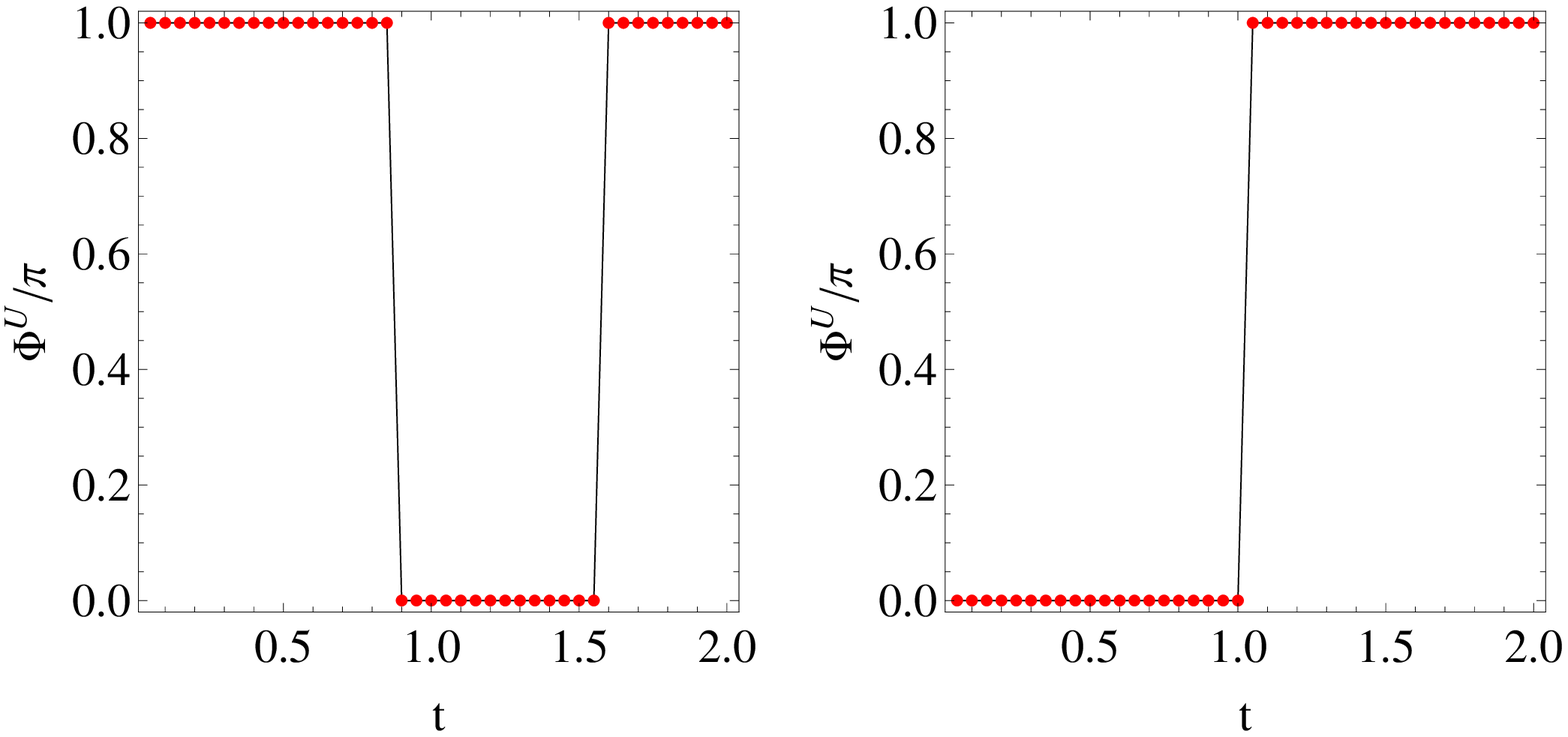}
\caption{The Uhlmann phase $\Phi^U$ as a function of time for the Creutz ladder of Eq.(\ref{eq-CL}). Left panel: $m$ changes from $0.6$ to $1.6$. Right panel: $m$ changes from $1.6$ to $0.6$. The inial temperature $T=0.4$ and $\theta=0.3\pi$ for both panels.}
\label{fig-U-m}
\end{figure}

At finite temperature, we can study the time evolving of the Uhlmann phases instead of the Berry phases under the quench process.
Before we get into any details, we will first briefly review the topological properties of Creutz ladder in thermal equilibrium states.

As before, the Hamiltonian of Creutz ladder can be written as $H=R_i\sigma_i$. For a two-band model, its two eigenvalues take very simple forms as $E_{1,2}=\pm R$, with $R=\sqrt{R_i^2}$. The corresponding Boltzmann weights of the two eigenstates are
\be
p_{1,2}=\frac{e^{\mp R/T}}{Z},\qquad Z=2\cosh(R/T)
\ee
The eigenstates $\ket{u_{1,2}}$ can be described by the two projection operators as
\be
P_{1,2}=\frac12(1\pm\hat{R}_i\sigma_i)
\ee
here $\hat{R_i}=R_i/R$. Then it is easy to find the density matrix as
\be
\rho=p_i P_i=\frac12\Big(1-\tanh(\frac{R}{T})\hat{R}_i\sigma_i\Big)
\ee
Substitute the above results to Eq.(\ref{eq-AU}), we find the Uhlmann connection as
\be
A^U&=&f(R)(P_1\p_k P_2+P_2\p_k P_1)\nonumber\\
&=&-\frac{i}2f(R)\epsilon_{abc}\hat{R}_a\p_k\hat{R}_b\sigma_c
\label{eq-A-R}
\ee
Here $f(R)=1-\frac{1}{\cosh(R/T)}$ and we have used $\hat{R}_i\p_k\hat{R}_i=0$.

\begin{figure}
\centering
\includegraphics[width=0.4\columnwidth]{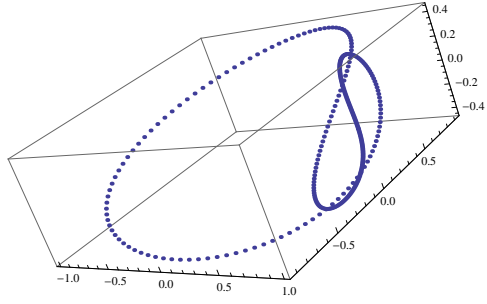}
\caption{The trajectory of $\hat{R}_i(k)$ for $k\in(-\pi,\pi)$ at the time of $t=1.2$.}
\label{fig-R-k}
\end{figure}

To compute the Uhlmann phase, we first consider the Uhlmann Wilson loop given by
\be
V=\mathcal{P}\exp\Big(\int_0^{2\pi} A^U d k\Big)
\label{eq-V}
\ee
For the Creutz ladder, the trajectory of $\hat{R}_i(k)$ is a circle on the $xz$ plane. In this case, we find that the vector of $\epsilon_{abc}\hat{R}_a\p_k\hat{R}_b$ is proportional to the $y$ axis. Therefore, the Uhlmann connection can be rewritten as
\be
A^U=-\frac{i}2f(R)(\hat{R}_3\p_k\hat{R}_1-\hat{R}_1\p_k\hat{R}_3)\,\sigma_2
\ee
which is proportional to a constant matrix. In this case, the path ordering becomes trivial and we find that
\be
V=\cos W-i\sigma_2\sin W,\quad W=\frac12\int_0^{2\pi} f(R)(\hat{R}_3\p_k\hat{R}_1-\hat{R}_1\p_k\hat{R}_3)dk
\ee
Now it is easy to find the Uhlmann phase as
\be
\Phi^U=\textrm{Tr}[\rho_0 V]=\cos W
\ee
Since $\cos W$ is a real number, $\Phi_U$ is always quantized for the Creutz ladder at thermal equilibrium. At very low $T$, we have $f(R)=1-1/\cosh(R/T)\approx1$. Thus the integral of $W$ becomes a form of wind number. For $|m|<1$, we find that $\Phi^U=W=\pi$ and for $|m|>1$, $\Phi^U=W=0$.  In other words, the Uhlmann phase recovers the result of Berry phase at very low $T$. On the other hand, for very large $T$, $f(R)\approx 0$. In this case, we always have $\Phi^U=W=0$. Clearly, in the topological region $|m|<1$, there is a critical temperature $T_c$, across which $\Phi^U$ jumps from $\pi$ to $0$.

Now we turn to the quench process. In Figure \ref{fig-U-m}, we plot the Uhlmann phase $\Phi^U$ of the Creutz ladder as a function of time after the quench. In the left panel, we start from the topological phase with $m=0.6$ and $T=0.4$ then suddenly change it to the trivial phase with $m=1.6$. One can see that $\phi^U=\pi$ initially and keep this quantized value for a while. At certain time, it abruptly jumps to $0$. At some later time, it jumps back to the non-trivial value $\pi$ again. In the right panel, we show the reversed case, starting from trivial phase then tune back to topological phase. Then we find that $\phi^U=0$ initially, but will jump to non-trivial value $\pi$ at certain point of time. In both cases, $\phi^U$ is still quantized to $0$ and $\pi$ after the quench. Although $T$ is fixed, the sudden change of parameter may cause topological transition in the sense of Uhlmann phase. This is in contrast to the Berry phase which is constant after the quench.

\begin{figure}
\centering
\includegraphics[width=0.7\columnwidth]{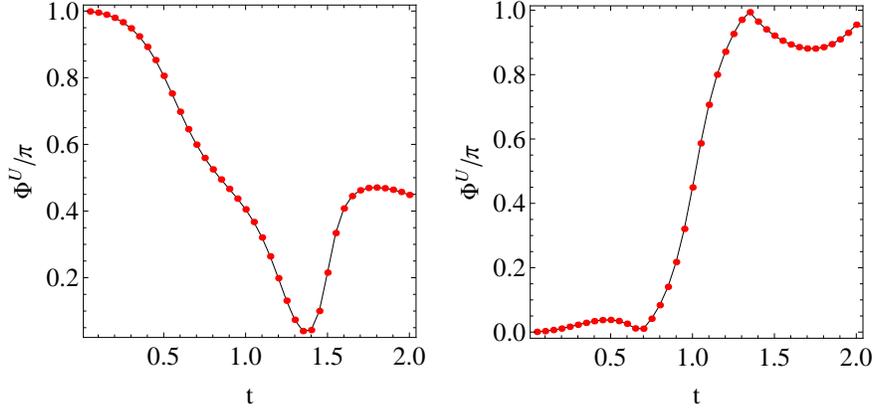}
\caption{The Uhlmann phase $\Phi^U$ as a function of time for the Creutz ladder of Eq.(\ref{eq-CLp}). Left panel: $(m,\gamma)$ changes from $(0.6,0)$ to $(1.6,0.3)$. Right panel: $m$ changes from $(1.6,0)$ to $(0.6,0.3)$. The initial temperature $T=0.4$ and $\theta=0.3\pi$ for both panels.}
\label{fig-U-gamma}
\end{figure}

One may be curious about why $\phi^U$ is still quantized after the quench. The initial Hamiltonian can be described by vectors $R^0_i(k)$.  After the quench, these vectors will be rotated around $R^1_i(k)$ to become a set of new vectors $R_i(k)$. Since the rotation axes also depend on $k$, the resulting curves of $\hat{R}_i(k)$ can not be located on the same plane. As an example, we show the curve of $\hat{R}_i(k)$ at $t=1.2$ in Figure \ref{fig-R-k}, which is clearly not a plane curve.
Although we can still substitute $\hat{R}_i(k)$ into the Eq.(\ref{eq-A-R}), the cross product $\epsilon_{abc}\hat{R}_a\p_k\hat{R}_b$ is not a constant vector any more. Therefore the product ordering in the Eq.(\ref{eq-V}) becomes quite non-trivial to calculate analytically. By numerical calculations, we found that the path ordering product $V$ is
\be
V\propto \exp\Big[i(c_2\sigma_2+c_3\sigma_3)\Big]
\label{eq-V-1}
\ee
with some real constants $c_2$ and $c_3$ depending on the evolving time. Since the initial density matrix is $\rho_0=(1+\sigma_1)/2$, we immediately see that Tr$(\rho_0V)$ is a real number and  $\Phi_U$ in this case is still quantized. Comparing with the computations of Berry phase under quench process, we suspect that it is still the symmetry about $x$ axis of the $R^0_i(k)$ and $R^1_i(k)$ curves that guarantees the $x$ component vanishes in the exponent of Eq.(\ref{eq-V-1}). This property in turn gives rise to a quantized $\Phi^U$.

\begin{figure}
\centering
\includegraphics[width=0.7\columnwidth]{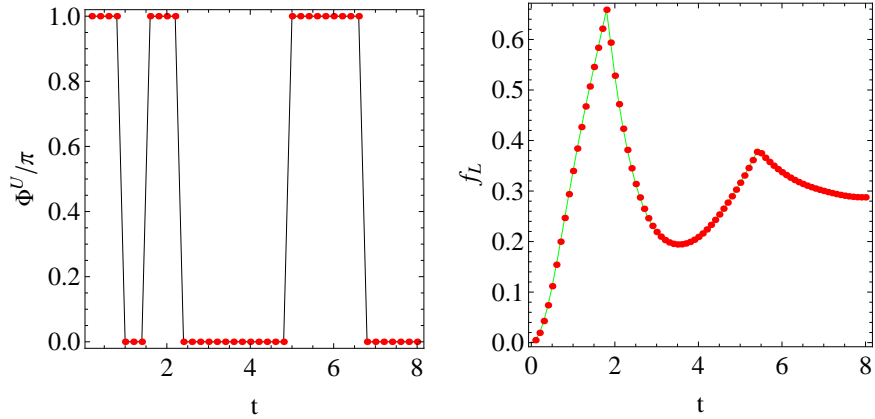}
\caption{The Uhlmann phase $\Phi^U$ (left panel) and Loschmidt rate function $f_L$ (right panel) as a function of time for Creutz ladder of Eq.(\ref{eq-CLp}). For both panels, $m$ changes from $0.6$ to $1.6$. Other parameters are $T=0.4$ and $\theta=0.3$.}
\label{fig-U-rate}
\end{figure}

To break the symmetry of $R^0_i(k)$ curve, we can still consider the modified Creutz ladder of Eq.(\ref{eq-CLp}) with an extra term $\gamma\sigma_2$. The Uhlmann phase of this model under the quench process is plotted in Figure \ref{fig-U-gamma}. In the left panel, the parameter change is chosen as $(m,\gamma)=(0.6,0)\to(1.6,0.3)$. In the right panel, we have $(m,\gamma)=(1.6,0)\to(0.6,0.3)$. In both cases, $\Phi^U$ deviates from the quantized values and keeps fluctuating with time. Only at certain points of time, $\phi^U$ can reach the quantized values again. But since there is no chiral symmetry in the modified Creutz ladder, these values have no clear topological meaning.

The abrupt jumps in the Figure \ref{fig-U-m} signal the topological transition under the quench process. One may wonder if this transition is related to the dynamical quantum phase transition. The dynamical quantum phase transition is associated with the Loschmidt amplitude and rate function defined as follows
\be
f_{\textrm{L}}=-\lim_{N\to\infty}\frac1N\sum_k\ln|\mathcal{G}(t)|^2,\quad
\mathcal{G}(t)=\textrm{Tr}\Big[\rho_k(0)e^{-iH_1 t}\Big]
\ee
Here $N$ is the system size and $\rho_k(0)$ is the density matrix at momentum $k$ at initial time. We compare the time dependence of Uhlamnn phase and rate function under the quench process in Figure \ref{fig-U-rate}. In this figure, we allow the system to evolve for a longer time up to $t=8$. One can see the time points of jumps in the $\Phi^U$ curve are closed to the peaks in the $f_{\textrm{L}}(t)$ curve. But the precise location of the jumps and the peaks cannot match to each other. Thus we cannot establish the exact relation between the Uhlamnn phase and the Loschmidt rate function.

\section{conclusion}
\label{sec-con}

In this paper, we have shown that the Berry phase of 1D topological models is constant under the quench process if the symmetry of the model is preserved. The time evolution after the quench can be visualized as a closed loop on a unit sphere deforming its shape with time. It is found that the value of Berry phase can be converted to the area of the surface bounded by this closed loop. Based on this observation, we have analytically shown that the symmetry of the model dictates that the Berry phase must be a constant. At finite $T$, we found that the Uhlamnn phase is always quantized to two possible values after the quench when the symmetry of the model is intact. As time flows, the system jumps between the topological non-trivial and trivial phases according to sudden changes of the Uhlamnn phase. If a term that breaks the chiral symmetry is included, then both the Berry and Uhlmann phases deviate from the quantized values. But the topological meaning of these geometric phases is not clear in this case.

\begin{acknowledgments}
Y. H. was supported by the Natural Science Foundation of China under Grant No. 11874272.
\end{acknowledgments}

%\bibliographystyle{apsrev}
%\bibliography{Reference}

\end{document}